\documentclass[sigconf]{acmart}

\AtBeginDocument{%
  \providecommand\BibTeX{{%
    \normalfont B\kern-0.5em{\scshape i\kern-0.25em b}\kern-0.8em\TeX}}}

\setcopyright{none}
\renewcommand\footnotetextcopyrightpermission[1]{}
\begin{document}

\title{Using Real-world Bug Bounty Programs in Secure Coding Course: Experience Report}

\author{Kamil Malinka}
\orcid{0000-0002-9009-2193}
\affiliation{%
  \institution{Brno University of Technology}
  \streetaddress{Božetěchova 2}
  \city{Brno} 
  \country{Czech Republic}
  \postcode{61200}  
}
\email{malinka@fit.vut.cz}

\author{Anton Firc}
\orcid{0000-0002-4717-1910}
\affiliation{%
  \institution{Brno University of Technology}
  \streetaddress{Božetěchova 2}
  \city{Brno} 
  \country{Czech Republic}
  \postcode{61200}  
}
\email{ifirc@fit.vut.cz}

\author{Pavel Loutocký}
\orcid{0000-0002-4965-1467}
\affiliation{%
  \institution{Masaryk University}
  \streetaddress{Botanická 554/68a}
  \city{Brno} 
  \country{Czech Republic}
  \postcode{60200}  
}
\email{Pavel.Loutocky@law.muni.cz}

\author{Jakub Vostoupal}
\orcid{0000-0002-1669-9931}
\affiliation{%
  \institution{Masaryk University}
  \streetaddress{Botanická 554/68a}
  \city{Brno} 
  \country{Czech Republic}
  \postcode{60200}   
}
\email{Jakub.Vostoupal@law.muni.cz}

\author{Andrej Krištofík}
\orcid{0000-0001-8150-0362}
\affiliation{%
  \institution{Masaryk University}
  \streetaddress{Botanická 554/68a}
  \city{Brno} 
  \country{Czech Republic}
  \postcode{60200}    
}
\email{Andrej.Kristofik@law.muni.cz}

\author{František Kasl}
\orcid{0000-0001-6675-9528}
\affiliation{%
  \institution{Masaryk University}
  \streetaddress{Botanická 554/68a}
  \city{Brno} 
  \country{Czech Republic}
  \postcode{60200}    
}
\email{Frantisek.Kasl@muni.cz}

\renewcommand{\shortauthors}{Malinka et al.}

\begin{abstract}
To keep up with the growing number of cyber-attacks and associated threats, there is an ever-increasing demand for cybersecurity professionals and new methods and technologies. Training new cybersecurity professionals is a challenging task due to the broad scope of the area. One particular field where there is a shortage of experts is Ethical Hacking. Due to its complexity, it often faces educational constraints. Recognizing these challenges, we propose a solution: integrating a real-world bug bounty programme into cybersecurity curriculum. This innovative approach aims to fill the gap in practical cybersecurity education and also brings additional positive benefits.

To evaluate our idea, we include the proposed solution to a secure coding course for IT-oriented faculty. 
We let students choose to participate in a bug bounty programme as an option for the semester assignment in a secure coding course. We then collected responses from the students to evaluate the outcomes (improved skills, reported vulnerabilities, a better relationship with security, etc.). Evaluation of the assignment showed that students enjoyed solving such real-world problems, could find real vulnerabilities, and that it helped raise their skills and cybersecurity awareness. Participation in real bug bounty programmes also positively affects the security level of the tested products. We also discuss the potential risks of this approach and how to mitigate them.

\end{abstract}


\keywords{University Education, Bug Bounty, Cybersecurity Specialization, Secure coding, Course Assignment, Experience Report}

\maketitle
\pagestyle{plain}
\section{Introduction}

The IT threat landscape continuously increases, resulting in a vastly increased demand for cybersecurity experts. However, the lack of cybersecurity experts is a long-known problem. One of the solutions to how we tried to tackle this problem is to use crowdsourcing solutions to share intelligence. An example of such efforts is bug bounty programs (BBPs). 

Bug Bounty Programs should be considered indispensable tools promoting responsible vulnerability disclosure~\cite{hata_understanding_2017}. These programs do not rely only on altruistic and randomly encountered ethical hackers. They incentivize them with rewards for reporting relevant cybersecurity vulnerabilities~\cite{hata_understanding_2017}. Vendors typically announce these specific competitions to external stakeholders, ranging from the general public to researchers and security-oriented companies~\cite{ruohonen_bug_2018}.

In the case of a bug bounty program, which is one of the branches of so-called ethical hacking,  they are coordinated to some extent by the entity that has an interest in discovering vulnerabilities in a given system (not necessarily its own). We find it essential to coordinate related activities within these approaches, not only in relation to the vulnerability testing itself but especially apriori at the level of education within specialized study programs to implement and introduce study projects that would lead to the use of bug bounty-related approaches as part of student motivation.

Thus, cybersecurity education can help address the shortage of cybersecurity professionals by introducing students to ethical hacking and allowing them to participate in real-world BBPs.

This cooperation between academia and real-world business yields several key benefits for both parties, fostering much-needed synergy. This includes endowing students with valuable real-life experiences that test their technical knowledge and cultivate non-technical skills, such as effective communication and report filing~\cite{radziwill_ethics_2015}.
Moreover, these experiences benefit students’ personal development and enhance their employability, leveraging the prestige associated with successful bounty hunts~\cite{hata_understanding_2017}.

We further argue that the positive effects of introducing BBPs into the cybersecurity curriculum are not limited only to the development of hard and soft skills but hold a much more profitable potential by instilling in students the principles of ethical hacking and cultivating an appreciation for the benefits of both BBPs and ethical hacking as a whole, even outside the course work~\cite{hartley_ethical_2015}.

In the context of this paper, we evaluate specific approaches and experiences with implementing bug bounties in the curriculum, the practical benefits for students, and the impact on their knowledge and skill base. It is also crucial to evaluate whether students of a particular IT program can solve specific tasks and requirements of bug bounty programs. Based on our teaching experience, we know that students like real-world problems and appreciate real-world examples. However, we were interested to see how they would cope with this type of problem, as it represents quite a big leap from conventional school problems.

To experimentally verify our proposal, we let students voluntarily select participating in a BBP as a semester project in a secure coding course. The success or failure of the search for the vulnerability was not reflected in the final grade. We evaluated only the process, used tools, and final report. We also organized an extra lecture for those interested in this project area to give students a basic orientation in the field of ethical hacking.

After completing the assignments, we surveyed the students to find out how they liked the possibility of being involved in solving real-world problems. We focus on three areas of questions: technical parameters, student self-evaluation, and project evaluation. Additionally, we examined how many vulnerabilities will be reported by BBP participants with no prior knowledge (students) and how the students perceive such an assignment (entertainment, skills, risks).

\paragraph{\textbf{Contributions.}}

The main contributions of this paper may be stated as follows:
\begin{itemize}
    \item We proposed an innovative way to improve the learning of IT and cybersecurity professionals that leverage real bug bounty programs in the curriculum.
    \item We have experimentally implemented and evaluated the proposed idea in one run of the university course.
    \item We discuss the pedagogical implications of the proposed approach and have shown, among other things, a positive impact on learning and that students can solve this type of task successfully.
\end{itemize}

\section{Motivation}

No IT solution is free of bugs and vulnerabilities~\cite{zetter_information_2001}, and finding and patching vulnerabilities is an iterative process that is both financially and human capital intensive~\cite{kinis_responsible_2018}. Letting external entities  (i.e. cybersecurity researchers, testers, enthusiasts or hackers) search for and responsibly disclose cybersecurity vulnerabilities can thus be an effective security tool to mitigate such vulnerabilities.

Such external help can be very welcome (or even needed) from the vendor's side~\cite{ruohonen_bug_2018}. A specific approach to gathering these external entities is the introduction of the already mentioned BBPs, where the system or device is subjected to "planned" attacks at the request of the vendor and the ethical hackers are then rewarded for reporting found vulnerabilities ~\cite{riggs_i_2021}. 
It is then essential to ensure that the relevant activities are legal not only from the perspective of the vendor but also from the perspective of the ethical hackers themselves so that they do not have to worry about unwanted sanctions (which would limit or even eliminate the motivation for conducting such actions~\cite{weulen_kranenbarg_dont_2018}). Nevertheless, it is essential to emphasize that many actors involved in the BBPs’ procedures (the notifying ethical hacker, the vendor, or third parties and coordinators) often have incompatible goals and interests~\cite{malladi_bug_2020}. The potential conflicts between these actors may seriously hinder the motivational aspect, and it is, therefore, crucial to mitigate this risk by thorough setup of the conditions for the announcement (and the whole of the procedure) under a bug bounty program~\cite{walshe_empirical_2020}. 

However, it has to be stated that participation and hands-on experiences in the context of these programs are not often encountered by students (not only) of higher education programs, although they can be of great benefit to them~\cite{hartley_ethical_2015, radziwill_ethics_2015, trabelsi_ethical_2016}. In our experience, which we present within this paper, this is a benefit not only in terms of gaining actual awareness and concrete, hands-on experience with such cyber security vulnerability detection but especially in terms of gaining superior related skills that appropriately further shape the specialized profile of the students themselves. 

In such regard, the cooperation between the educational institutions and the vendors has its own significance, as it may satisfy the vendors’ need for such expert labour, which is currently lacking in the European market~\cite{noauthor_urgency_2022}. 
On the other hand, this cooperation allows students to gain hands-on experience and help educators better prepare them for real scenarios.
Making the courses more interesting by utilising a real-world exercise is also an element that should not be overlooked~\cite{hartley_ethical_2015, radziwill_ethics_2015, trabelsi_ethical_2016}. Gaining actual awareness and concrete hands-on experience with such cyber security vulnerability detection, and especially gaining superior related skills~\cite{hartley_ethical_2015}  help further shape the students' specialised profile.

We must also consider another factor - the time allotment structure of the curriculum. 
The curriculum of security-oriented courses and specializations is  often broad and covers only the basics of all the areas concerned. 
Students are often taught the basics as a part of general security-oriented courses, but it is undesirable to neglect other areas such as cryptography, authentication, or IT security management in favour of ethical hacking. The education in the area of ethical hacking is thus often minimal and left to the students themselves. The situation is even more problematic in the area of IT students not focused on security. Although they should also have the basics of cybersecurity, they have even less teaching space. Also, incorporating a real-world problem into teaching brings with it a number of challenges in addition to the benefits mentioned above - greater difficulty may prevent students from successfully solving, ensuring assignment consistency for a fair assessment, or emphasis on practical knowledge of a wide range of tools.

\section{Related work}

The use of bug bounty programs for controlled vulnerability discovery is very common in the case of large companies (e.g. Apple\footnote{\url{https://security.apple.com/bounty/}}, Samsung\footnote{\url{https://security.samsungmobile.com/rewardsProgram.smsb}} or Microsoft\footnote{\url{https://www.microsoft.com/en-us/msrc/bounty}}). Such programs usually have a graduated range of rewards according to the vulnerability discovered in the general terms of the program. It is thus evident from practice that the introduction of bug bounty programmes brings substantial benefits (both financial and security), which are all the more evident in conjunction with specialised service providers, which usually act as intermediaries in the processing of the programme. These include companies such as BugCrowd\footnote{\url{https://www.bugcrowd.com/products/platform/}} and HackerONE\footnote{\url{https://www.hackerone.com/}}. Programs offered by these services were used in the 2020 quantitative study conducted by Walshe and Simpson~\cite{walshe_empirical_2020} that has demonstrated how a well-deployed program could, in financial terms, substitute for two full-time experts. 

In the context of education, however, the problematic implementation of practical knowledge related to bug bounty programmes into university teaching is evident~\cite{hartley_ethical_2015, radziwill_ethics_2015, trabelsi_ethical_2016}\footnote{On the other hand, there is a relatively large number of specifically targeted courses, especially in the online environment. See e.g. here: \url{https://securitytrails.com/blog/popular-bug-bounty-courses} or here: \url{https://www.classcentral.com/report/best-bug-bounty-courses/}}. There is a generally noticeable orientation towards broader educational focuses that focus on, for example, cybersecurity specialists; thus, a specific targeting on bug bounty-related skills is sporadic\footnote{This is also evident from the report within the SPARTA project - \url{https://ec.europa.eu/research/participants/documents/downloadPublic?documentIds=080166e5d212c432&appId=PPGMS}, e.g.. p. 53 or 56.}, even though it may be deemed a crucial part of the cybersecurity professionals’ skillset ~\cite{pashel_teaching_2006}. In the context of cybersecurity education, this was emphasized already by Greene ~\cite{greene_training_2004} and also in the research of Logan and Clarkson ~\cite{logan_teaching_2005} and Pashel ~\cite{pashel_teaching_2006}, and studied further by Trabelski (and others) ~\cite{trabelsi_hands-lab_2011, trabelsi_switchs_2012, trabelsi_using_2013, trabelsi_enhancing_2014, trabelsi_ethical_2016} and Hartley (and others) ~\cite{hartley_ethical_2015, hartley_ethical_2017}.

We aim to showcase the benefits and insights gained from integrating bug bounty programs into our curriculum, echoing Hartley's findings on the significant impact of hands-on experience in cybersecurity education~\cite{hartley_ethical_2015}. 

We advocate that a proper curriculum design should give the students hands-on experience and provide them with the necessary soft skills and knowledge outside of the tech domain. This part of the curriculum should also focus on the legal and ethical aspects of hacking, as Trablesi et al.~\cite{trabelsi_ethical_2016} have shown in their research that there is a potential for malicious activity done by the students.

\section{Proposed solution}

This section presents a possible solution to the above-mentioned problems by combining education and real-world ethical hacking. Our innovative approach aims to fill the practical cybersecurity education gap and bring other positive benefits.

To evaluate our idea, we implemented the proposed solution to a course for IT-oriented faculty. 
We let students choose to participate in a bug bounty programme as an option for the semester project in a secure coding course. We then collected responses from the students to evaluate the outcomes.
In addition to the technical question of whether students without prior expertise in computer security would even be able to successfully solve a BBP and what procedures and tools they would use, we also investigated how students evaluate this type of project and what the implications are for teaching this topic.







It has to be said that there are several risks in letting the students participate in BBP. Firstly, the students may not be able to identify any vulnerability. While this may be a problem for the BBP itself, the project focuses on the process and educational benefits. Shortly said, even failure to identify any vulnerability is considered positive if the students improve their skills and broaden their horizons. Secondly, the students may violate one of the rules of the BBP. 

To mitigate the risks, we provide a proper introduction to ethical hacking and BBPs in the form of a lecture delivered by an expert in the field. An integral component of the course involves teaching students the necessary skills of the ethical hacker~\cite{hartley_ethical_2015}.
These encompass technical proficiency and a fundamental understanding of the relevant legal framework and the ability to navigate the legal specifics, rules, and methodologies of the particular BBP~\cite{hartley_ethical_2017}.

This ensures that their actions are conducted within legal boundaries, minimizing the risk of causing undue harm and facing subsequent consequences
~\cite{hartley_ethical_2017}.

We propose specifically dedicating one of the introductory lessons to the legal issues, where students are acquainted with locating the specific rules of a given BBP and the potential consequences of overstepping these rules or targeting services that fall outside the defined boundaries. This approach aims to establish a secure foundation for both students and educators, fostering a safe learning environment. Additionally, it assists in orienting them toward ethical hacking methodologies and ensures a responsible and lawful engagement with bug bounty initiatives~\cite{trabelsi_ethical_2016}.


\section{Course project description}

The experiment was conducted as part of an optional university course focused on secure coding. The course is regularly taught in computer security specialization at the IT-oriented faculty. The course is designed for Master's students, introduces the basic principles of secure coding, and explains the general principles of vulnerabilities and defences against them. It covers multiple areas, such as basic vulnerabilities of compiled languages, memory protection mechanisms, input validation, static and dynamic analysis, and more, but in general, the course is not focused on ethical hacking.

An important part of the course is the semester project, for which students can get almost half of all points counted in the final grade. Students have two months to finish the project. Students are expected to work independently on a selected topic that falls within the course's content area. They are expected to study the relevant materials, research the chosen area, and possibly implement selected solutions. 

The result is a technical article in the selected area of at least six pages in double-column IEEE format (for implementation, the output may differ by agreement). An integral part of the solution is the oral presentation of the whole work, which takes place in the course seminars. The goal is for students to demonstrate both hard and soft skills, as this combination is expected of future cybersecurity professionals.

Students choose their own topic, but the course teacher must always approve it. They choose from 3 types of projects: \textit{tutorial}, \textit{HW or SW implementation}, and \textit{original work}.

In the \textit{Tutorial}, students have to study a selected topic in depth and write a short tutorial or overview study with the structure of a scientific paper. Their own opinion and analysis are welcome.
In the \textit{Implementation}, they choose an algorithm and a platform on which they then implement the algorithm with respect to a specific set of goals (speed, security, and/or performance). The goal is to produce a working code and write a paper explaining the implementation.
Within the last area, \textit{Original work}, they can present their own activities if they work on an interesting, relevant topic from the secure coding area: a new method, algorithm, implementation technique, optimization of existing solutions, etc. The form of the solution is again a paper describing the work.

As part of the experiment, we have added a new voluntary area - \textit{Bug Bounty challenge}. The challenge expected active participation in the selected bug bounty program and an attempt to find a real vulnerability. A list of suitable BBPs was provided, but self-selection was also allowed to enable students to pick the most technologically suitable BBP. The success or failure was not reflected in the final grade. We evaluated only the process, used tools, and final report. Thus, if one failed to find something specific, it did not compromise the passing of the course in any way. In case of success, correct reporting of the vulnerability according to the processes of the selected BBP was mandatory. Students were repeatedly reminded of the need to strictly follow the competition's rules.

For those interested in ethical hacking, we organized an additional lecture to provide a basic orientation in the field. Delivered by an expert with over five years of experience, the lecture covered the motivations for ethical hacking and its appropriate applications. It introduced the primary methodologies and tools for web, infrastructure, and mobile environments. We also guided students through relevant learning platforms and Bug Bounty Programs (BBPs). Finally, we detailed how to report vulnerabilities, comply with BBP guidelines, and understand the legal framework.


After completing all the preparatory phases, students had 7 weeks to complete the entire project. After submitting and evaluating the entire project, students were contacted and asked to complete a questionnaire. In addition to the questionnaires, the evaluation included an analysis of the submitted project reports. 

The course's general learning outcomes are designed so that students can gain knowledge in the chosen security area, acquire the ability to write a professional text and acquire the skills to present professional content.
The specific learning outcomes for a BBS-type project are designed to enable students to:
\begin{itemize}
\item understand the skills and qualifications to be an ethical hacker,
\item demonstrate the knowledge of information gathering, testing, and ethically hacking a system,
\item learn about the different tools and techniques that hackers—including ethical hackers—employ,
\item use selected tools in hacking,
\item understand cybersecurity laws and the consequences for breaking those laws.
\end{itemize}

The course strengthens the following working life skills: presentation skills, creativity, problem-solving skills, and information and communication technology skills.

\section{Experiment Design}
The experiment began with the preparation of an extended project assignment and the preparation of questionnaires. As part of the publication of the assignment, students were informed in advance that this project is part of an experiment and were asked to consent to the processing of information about the experiment, including their use for the final publication.

After we published the assignment, students could attend a bonus lecture and work throughout the project. 

To evaluate the experiment, we used a combination of quantitative and qualitative methods. After the project was handed in, we sent a questionnaire to all participants, asking them additional questions. Further information was obtained through a detailed analysis of the submitted project reports, which included a description of the project solution, technical details, approach description, etc. Analysis of the results was carried out by team members who are also responsible for teaching and assessing the course.

As part of the analysis of the project reports, we focused on collecting information relevant to the responsible reporting process. However, we also discuss other interesting observations.

Within the questionnaires, we focused on 3 main areas. The results are intended to provide a better understanding of the impact of each project parameter. The first batch of questions was focused on the actual work done on the project. These questions were intended to discover the student's previous experience, the influence of his/her knowledge on choosing a particular BBP, the vulnerability search strategy, the tools used, the methodology, and many others. 


The second area focused on self-evaluation, where we asked students to evaluate their skills and knowledge before and after the project.
The last area concerned the students' evaluation of the project, where we were interested in the fun and usefulness of the project, the perception of risks, etc.

Data were collected and processed anonymously based on the consent of the participants in the experiment. According to the university's internal rules, the course supervisor and department head approved the whole experiment.

\section{Results}

A total of 38 students signed up for the course project, and a total of 31 students successfully completed it (which is comparable to past years). 19 students chose the experimental bug bounty challenge (BBCh). 13 of whom had BBCh assignments successfully completed the project (12 of them filled out the follow-up survey).

We did not observe a significant deviation from other project types in the project evaluation. The average project score of other types was 39 points out of 49 possible. The average BBCh project score was 40 points. Thus, in terms of difficulty, we rate all types of assignments as comparable.

\subsection{Student's work on the assignment}

In the first step, we were interested in the BBP choice. Five students chose the T-Mobile bug bounty program because they wanted to pursue the program available in their native language. Three chose HackerOne - specifically Shopify's Bug Bounty, Epic Games, and Boozt Fashion AB.
Other programs were covered by only one of the students: TryHackMe, HackTheBox, Hacker101, PicoCTF, Coinbase, Moneta, Microsoft, and Hacktrophy. 

Some students tried to participate in multiple programs. One of the students even used the obtained knowledge and did a back check of his code and found many errors.

The previous experience did not play any significant role in selecting this assignment. On a scale of 1 - 5 (no impact - high impact), the average score was 3.16 with a standard deviation $\sigma$ of 1.4. In most cases, the reason for selecting a specific BBP was prior experience with and knowledge of the given system. In addition, the majority of the selected BBPs were focused on the security of web applications. Some students even reported that they perceived the web applications as the easiest, thus suitable for beginners. 

The selection of BBP based on these factors is expected. Moreover, interacting with a familiar system should make it easier for inexperienced participants to find new vulnerabilities. 

In most cases, the strategy for identifying the vulnerabilities involved following checklists, guides, or tutorials (such as OWASP WSTG\footnote{\url{https://owasp.org/www-project-web-security-testing-guide/}}), testing for known vulnerabilities, or searching for vulnerabilities based on prior personal experience. 


Time-wise, most students have spent 15-30 hours with theoretical preparation. Three students have spent less than 10 hours, with a minimum of two. Three students have spent over 30 hours, with a maximum of 80. The vulnerability identification took 15-30 hours for most of the students. Only one student took less than 15 hours, and two more than 30 hours. Thus, most students were within the recommended time frame set for this project (40-60 hours).

The majority of students utilised the materials we suggested for their additional learning.
The most used education resource was PortSwigger Academy, as reported by 8 students, closely followed by video tutorials, reported by 5 students. The summary of used learning resources is shown in \autoref{tab:learningResources}.

\begin{table}
    \centering
    \caption{The overview of used learning resources.}
    \begin{tabular}{@{}lc@{}}
         \toprule
         \emph{Resource} & \emph{Times reported} \\ \midrule
         PortSwigger academy & 8 \\
         Videos & 5 \\
         TryHackMe & 3 \\
         OWASP & 2 \\
         Online blogs & 2 \\
         Hacker101 & 2 \\
         School lectures & 1 \\
         Scientific literature & 1 \\
         \bottomrule
    \end{tabular}
    \label{tab:learningResources}
\end{table}





Students used many tools to tackle the project, including BurpSuite, Wappalyzer, Kali Linux, Nmap, ffuf, Metasploit, and others. This reflects the variability of the BBP, as different implementations of the tested systems require different analytical tools.


We were most interested in seeing if any real vulnerabilities were found. Two of the students have found some vulnerabilities. The first one discovered an Insecure direct object reference (IDOR) vulnerability but did not report it, as the tested platform did not have a proper BBP. The second student discovered two vulnerabilities (CSRF and incorrect validation of redirect link after login) that he reported to the BBP, but at the time of submitting the paper, no response had yet been received. 
Furthermore, students found a small number of cases of non-standard behaviour, e.g., the existence and availability of non-actual pages, error messages available to users, etc. It is unclear whether or not these instances could be used to mount a successful attack. However, they still violate best practices and should potentially be fixed.

\begin{table*}
    \centering
    \caption{Personal likings of the project as reported by students. The responses ranged from 1 - 5, where 1 means the worst (negative) and 5 means the best (positive). $\sigma$ denotes the standard deviation.}
    \begin{tabular}{@{}lll@{}}
        \toprule
         Question & Mean & $\sigma$  \\ \midrule
         How much would you like to see this type of project incorporated into regular teaching? & 4 & 0.85 \\
         Is the project content beneficial even if the participant has a career path outside of cybersecurity? & 3.91 & 0.79 \\
         How do you feel about working in a real environment on real products? & 3.92 & 0.99 \\
         How important is the project's social impact to you (helping to improve real safety) & 3.42 & 1.08 \\
         \bottomrule
    \end{tabular}
    \label{tab:personalPreferences}
\end{table*}

\subsection{Self-evaluation of educational impacts}

\begin{figure}[t]
    \centering
    \includegraphics[width=\linewidth]{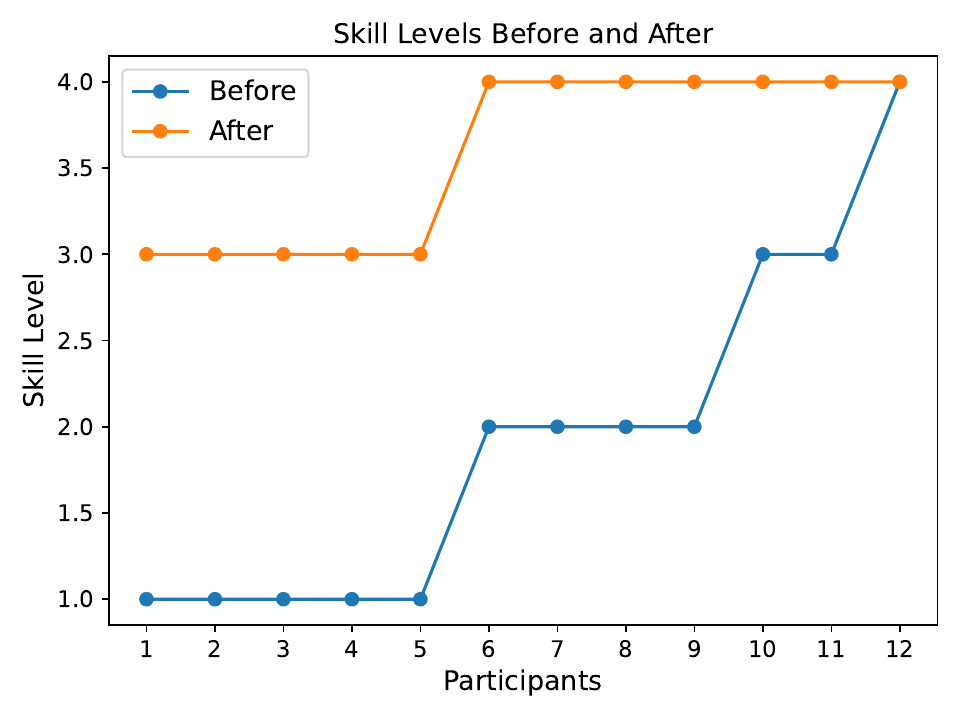}
    \caption{Orientation of participants on the topic of ethical hacking before and after the project. Participants were ordered by skill level before the project.}
    \Description{A line plot of the participants' orientation on ethical hacking before and after the project. Participants were ordered by skill level before the project.}
    \label{fig:self-eval}
\end{figure}

Before the project, participants in the experiment rated their understanding of ethical hacking as 1.92 on average; this increased to 3.58 at the end. \autoref{fig:self-eval} shows that in most cases, the students increased their skill in ethical hacking by two points.


We also assessed participants' prior knowledge of ethical hacking before the project. Nearly half had only theoretical understanding, while the rest had minimal practical experience from another school project. One participant had completed multiple TryHackMe courses and attended sessions on specific vulnerabilities at Burp Suite Academy.

After the end of the project, students reported good orientation in the topic, practical understanding of the different phases of penetration testing, understanding of the attacker's point of view and abilities, and practical knowledge of tools for ethical hacking.


41,7 \% of the students reported that they see their future career in cybersecurity before starting the assignment. After completing the assignment, it changed to 50\%.

We also wondered whether the students planned to participate in other bug bounty projects after the end of this project. Only two (17\%) do not plan to do so, six students (50\%) are considering it if there is enough time, and four (33\%) firmly plan to continue.

Ultimately, students found the knowledge gained beneficial, even if they pursued careers outside of cybersecurity.

\subsection{Self-assessment of the project}

Most students rated the experimental project as very beneficial, primarily due to the great practical overlap and the ability to work on real problems from practice. They also positively evaluated the relatively high degree of freedom of the project and, paradoxically, the need for self-study to a greater than usual extent, as the project allowed them to get an assessment in an area that interested them. One student summed it up this way: "\textit{If this project is chosen by a person who is interested and fond of this field, it is the most useful and interesting project at the school at all.}"

Next, we asked about the enjoyability of the project as a whole and how it compares to conventional projects with a fixed specification. On a scale of 1 - 5 (not enjoyable at all - it's extremely enjoyable), the average score was 3.81 with a standard deviation $\sigma$ of 0.87.
On a scale of 1 - 5 (significantly more boring than a regular project - significantly more entertaining than a regular project), the average score was 4.16 with a standard deviation $\sigma$ of 0.72.

When explaining the reasons for the assessment, students primarily mentioned
learning new skills, having the freedom to decide how to complete the assignment, and their involvement in trying to solve real-world problems. 
Despite the high values, almost a third of respondents mentioned time pressure due to the demands of other courses as well as the freedom of the course.

We were also interested in participants' views on questions regarding their personal likings of the project as shown in \autoref{tab:personalPreferences}.




Students generally rated the project as having a high practicality and educational impact. 
Negatives included the stress of breaking BBP rules or the frustration of not finding any vulnerabilities.


\section{Discussion and limitations}

The findings of this study have provided several noteworthy insights regarding the implementation and acceptance of our project-oriented approach to cybersecurity education. Firstly, to our surprise, students could identify vulnerabilities, which unequivocally endorses the efficacy of our pedagogical strategy. 

Secondly, the positive feedback received from students about this type of real-world project learning is particularly encouraging. Their enthusiasm underscores the relevance and engagement of hands-on, project-based learning in the conditions of the real world.

Moreover, we consider it of paramount importance that the project was deemed beneficial by those participants who do not intend to pursue a career in cybersecurity. The fact that students outside the field perceive the adversarial perspective as advantageous suggests that the skills and thought processes developed through this project have a wide-reaching impact, extending beyond the immediate domain of cybersecurity.
Ethical considerations in teaching hacking techniques within an educational setting have been a topic of some debate. However, various scholarly works have discussed this issue, and our stance aligns with the perspective that such education is legitimate ~\cite{hartley_ethical_2015}. By introducing IT students to the concept of thinking like an attacker, we foster a critical mindset for developing more secure systems. This cognitive shift is essential for producing programmers who are adept at anticipating and mitigating potential security threats.


Finally, we would like to summarize the main lessons learned.
It proved essential to allow freedom in the students' choice of technology by selecting a suitable BBS. Ethical hacking assumes a detailed knowledge of the technology, and it is usually not within the capabilities of the course to deliver this knowledge. Thus, students have to use their existing knowledge, which can be varied and diverse. To reduce the time commitment and increase the chances of success, it is thus advisable to let students work in a familiar environment.
We are also very positive about the feasibility of our approach, which was also helped by focusing the assessment on progression rather than finding vulnerabilities.
According to the feedback, we are considering how to appropriately integrate existing online courses on ethical hacking, which students widely used to gain detailed knowledge. 

We also considered the varying technical skills of students. Some needed more time than the average recommendation due to larger skill gaps, particularly for introductory education. However, this was accommodated by allocating sufficient credits for the project.








\subsection{Limitations}
We consider the limited dataset as a limitation, as the experiment was only conducted within one run of one course. However, we believe the results are sufficient for the initial opening of the debate and the data-driven presentation of our idea.
We are also planning further extensions and repeated runs of experiments.

\section{Conclusions}

Based on the positive feedback from students and valuable educational impacts, we plan to include BBP in future courses as well. We plan to shift the orientation towards government agencies, as such agencies often lack the resources to run their own programs or contract ethical hackers. Such a connection would not only be beneficial for students but also increase the security of the public sector.

\begin{acks}
This work was supported by the Brno University of Technology internal project FIT-S-23-8151.
\end{acks}

\bibliographystyle{ACM-Reference-Format}
\bibliography{sample-base}


\begin{thebibliography}{20}


\ifx \showCODEN    \undefined \def \showCODEN     #1{\unskip}     \fi
\ifx \showDOI      \undefined \def \showDOI       #1{#1}\fi
\ifx \showISBNx    \undefined \def \showISBNx     #1{\unskip}     \fi
\ifx \showISBNxiii \undefined \def \showISBNxiii  #1{\unskip}     \fi
\ifx \showISSN     \undefined \def \showISSN      #1{\unskip}     \fi
\ifx \showLCCN     \undefined \def \showLCCN      #1{\unskip}     \fi
\ifx \shownote     \undefined \def \shownote      #1{#1}          \fi
\ifx \showarticletitle \undefined \def \showarticletitle #1{#1}   \fi
\ifx \showURL      \undefined \def \showURL       {\relax}        \fi
\providecommand\bibfield[2]{#2}
\providecommand\bibinfo[2]{#2}
\providecommand\natexlab[1]{#1}
\providecommand\showeprint[2][]{arXiv:#2}

\bibitem[noa(2022)]%
        {noauthor_urgency_2022}
 \bibinfo{year}{2022}\natexlab{}.
\newblock \bibinfo{title}{The {Urgency} of {Tackling} {Europe}’s
  {Cybersecurity} {Skills} {Shortage}}.
\newblock
\newblock
\urldef\tempurl%
\url{https://blogs.microsoft.com/eupolicy/2022/03/23/the-urgency-of-tackling-europes-cybersecurity-skills-shortage/}
\showURL{%
\tempurl}


\bibitem[Greene(2004)]%
        {greene_training_2004}
\bibfield{author}{\bibinfo{person}{Tim Greene}.}
  \bibinfo{year}{2004}\natexlab{}.
\newblock \bibinfo{title}{Training {Ethical} {Hackers}: {Training} the
  {Enemy}?}
\newblock
\newblock
\urldef\tempurl%
\url{https://defcon.org/html/links/dc_press/archives/12/ebcvg_training_ethical_hackers.htm}
\showURL{%
\tempurl}


\bibitem[Hartley(2015)]%
        {hartley_ethical_2015}
\bibfield{author}{\bibinfo{person}{Regina Hartley}.}
  \bibinfo{year}{2015}\natexlab{}.
\newblock \showarticletitle{Ethical {Hacking} {Pedagogy}: {An} {Analysis} and
  {Overview} of {Teaching} {Students} to {Hack}}.
\newblock \bibinfo{journal}{\emph{Journal of International Technology and
  Information Management}}  \bibinfo{volume}{24} (\bibinfo{date}{Jan.}
  \bibinfo{year}{2015}), \bibinfo{pages}{95--104}.
\newblock
\urldef\tempurl%
\url{https://doi.org/10.58729/1941-6679.1055}
\showDOI{\tempurl}


\bibitem[Hartley et~al\mbox{.}(2017)]%
        {hartley_ethical_2017}
\bibfield{author}{\bibinfo{person}{Regina Hartley}, \bibinfo{person}{Dawn
  Medlin}, {and} \bibinfo{person}{Zach Houlik}.}
  \bibinfo{year}{2017}\natexlab{}.
\newblock \showarticletitle{Ethical {Hacking}: {Educating} {Future}
  {Cybersecurity} {Professionals}}.
\newblock \bibinfo{journal}{\emph{Information Systems \& Computing Academic
  Professionals: Proceedings of the EDSIG Conference}} (\bibinfo{year}{2017}),
  \bibinfo{pages}{1--10}.
\newblock


\bibitem[Hata et~al\mbox{.}(2017)]%
        {hata_understanding_2017}
\bibfield{author}{\bibinfo{person}{Hideaki Hata}, \bibinfo{person}{Mingyu Guo},
  {and} \bibinfo{person}{Muhammad Ali~Babar}.} \bibinfo{year}{2017}\natexlab{}.
\newblock \showarticletitle{Understanding the {Heterogeneity} of {Contributors}
  in {Bug} {Bounty} {Programs}}. In \bibinfo{booktitle}{\emph{2017 {ACM}/{IEEE}
  {International} {Symposium} on {Empirical} {Software} {Engineering} and
  {Measurement} ({ESEM})}}. \bibinfo{address}{Toronto, ON, Canada},
  \bibinfo{pages}{223--228}.
\newblock
\urldef\tempurl%
\url{https://doi.org/10.1109/ESEM.2017.34}
\showDOI{\tempurl}


\bibitem[Logan and Clarkson(2005)]%
        {logan_teaching_2005}
\bibfield{author}{\bibinfo{person}{Patricia~Y. Logan} {and}
  \bibinfo{person}{Allen Clarkson}.} \bibinfo{year}{2005}\natexlab{}.
\newblock \showarticletitle{Teaching students to hack: curriculum issues in
  information security}.
\newblock \bibinfo{journal}{\emph{ACM SIGCSE Bulletin}} \bibinfo{volume}{37},
  \bibinfo{number}{1} (\bibinfo{date}{Feb.} \bibinfo{year}{2005}),
  \bibinfo{pages}{157--161}.
\newblock
\showISSN{0097-8418}
\urldef\tempurl%
\url{https://doi.org/10.1145/1047124.1047405}
\showDOI{\tempurl}


\bibitem[Malladi and Subramanian(2020)]%
        {malladi_bug_2020}
\bibfield{author}{\bibinfo{person}{Suresh~S. Malladi} {and}
  \bibinfo{person}{Hemang~C. Subramanian}.} \bibinfo{year}{2020}\natexlab{}.
\newblock \showarticletitle{Bug {Bounty} {Programs} for {Cybersecurity}:
  {Practices}, {Issues}, and {Recommendations}}.
\newblock \bibinfo{journal}{\emph{IEEE Software}} \bibinfo{volume}{37},
  \bibinfo{number}{1} (\bibinfo{date}{Jan.} \bibinfo{year}{2020}),
  \bibinfo{pages}{31--39}.
\newblock
\showISSN{1937-4194}
\urldef\tempurl%
\url{https://doi.org/10.1109/MS.2018.2880508}
\showDOI{\tempurl}
\newblock
\shownote{Conference Name: IEEE Software}.


\bibitem[Pashel(2006)]%
        {pashel_teaching_2006}
\bibfield{author}{\bibinfo{person}{Brian~A. Pashel}.}
  \bibinfo{year}{2006}\natexlab{}.
\newblock \showarticletitle{Teaching students to hack: ethical implications in
  teaching students to hack at the university level}. In
  \bibinfo{booktitle}{\emph{Proceedings of the 3rd annual conference on
  {Information} security curriculum development}}. \bibinfo{publisher}{ACM},
  \bibinfo{address}{Kennesaw Georgia}, \bibinfo{pages}{197--200}.
\newblock
\showISBNx{978-1-59593-437-6}
\urldef\tempurl%
\url{https://doi.org/10.1145/1231047.1231088}
\showDOI{\tempurl}


\bibitem[Radziwill et~al\mbox{.}(2015)]%
        {radziwill_ethics_2015}
\bibfield{author}{\bibinfo{person}{Nicole Radziwill}, \bibinfo{person}{Jessica
  Romano}, \bibinfo{person}{Diane Shorter}, {and} \bibinfo{person}{Morgan
  Benton}.} \bibinfo{year}{2015}\natexlab{}.
\newblock \showarticletitle{The {Ethics} of {Hacking}: {Should} {It} {Be}
  {Taught}?}
\newblock \bibinfo{journal}{\emph{Software Quality Professional}}
  \bibinfo{volume}{18}, \bibinfo{number}{1} (\bibinfo{date}{Dec.}
  \bibinfo{year}{2015}).
\newblock
\urldef\tempurl%
\url{https://arxiv.org/abs/1512.02707}
\showURL{%
\tempurl}


\bibitem[Riggs(2021)]%
        {riggs_i_2021}
\bibfield{author}{\bibinfo{person}{Jacob Riggs}.}
  \bibinfo{year}{2021}\natexlab{}.
\newblock \bibinfo{title}{I hacked the {Dutch} government and all {I} got was
  this t-shirt}.
\newblock
\newblock
\urldef\tempurl%
\url{https://jacobriggs.io/blog/posts/i-hacked-the-dutch-government-and-all-i-got-was-this-t-shirt-24.html}
\showURL{%
\tempurl}


\bibitem[Ruohonen and Allodi(2018)]%
        {ruohonen_bug_2018}
\bibfield{author}{\bibinfo{person}{Jukka Ruohonen} {and} \bibinfo{person}{Luca
  Allodi}.} \bibinfo{year}{2018}\natexlab{}.
\newblock \showarticletitle{A {Bug} {Bounty} {Perspective} on the {Disclosure}
  of {Web} {Vulnerabilities}}.
\newblock \bibinfo{journal}{\emph{17th Annual Workshop on the Economics of
  Information Security, Innsbruck}} (\bibinfo{date}{May} \bibinfo{year}{2018}).
\newblock
\urldef\tempurl%
\url{http://arxiv.org/abs/1805.09850}
\showURL{%
\tempurl}
\newblock
\shownote{arXiv: 1805.09850}.


\bibitem[Trabelsi(2011)]%
        {trabelsi_hands-lab_2011}
\bibfield{author}{\bibinfo{person}{Zouheir Trabelsi}.}
  \bibinfo{year}{2011}\natexlab{}.
\newblock \showarticletitle{Hands-on lab exercises implementation of {DoS} and
  {MiM} attacks using {ARP} cache poisoning}. In
  \bibinfo{booktitle}{\emph{Proceedings of the 2011 {Information} {Security}
  {Curriculum} {Development} {Conference}}}. \bibinfo{publisher}{ACM},
  \bibinfo{address}{Kennesaw Georgia}, \bibinfo{pages}{74--83}.
\newblock
\showISBNx{978-1-4503-0812-0}
\urldef\tempurl%
\url{https://doi.org/10.1145/2047456.2047468}
\showDOI{\tempurl}


\bibitem[Trabelsi(2012)]%
        {trabelsi_switchs_2012}
\bibfield{author}{\bibinfo{person}{Zouheir Trabelsi}.}
  \bibinfo{year}{2012}\natexlab{}.
\newblock \showarticletitle{Switch's {CAM} table poisoning attack}. In
  \bibinfo{booktitle}{\emph{Computing {Education} 2012 - {Proceedings} of the
  14th {Australasian} {Computing} {Education} {Conference}}}
  \emph{(\bibinfo{series}{Conferences in {Research} and {Practice} in
  {Information} {Technology} {Series}})},
  \bibfield{editor}{\bibinfo{person}{Michael de~Raadt} {and}
  \bibinfo{person}{Angela Carbone}} (Eds.). \bibinfo{publisher}{Australian
  Computer Society}, \bibinfo{address}{Melbourne, Australia},
  \bibinfo{pages}{113--120}.
\newblock
\showISBNx{978-1-921770-04-3}
\urldef\tempurl%
\url{http://www.scopus.com/inward/record.url?scp=85014905333&partnerID=8YFLogxK}
\showURL{%
\tempurl}
\newblock
\shownote{Publisher: Australian Computer Society}.


\bibitem[Trabelsi(2014)]%
        {trabelsi_enhancing_2014}
\bibfield{author}{\bibinfo{person}{Zouheir Trabelsi}.}
  \bibinfo{year}{2014}\natexlab{}.
\newblock \showarticletitle{Enhancing the comprehension of network sniffing
  attack in information security education using a hands-on lab approach}. In
  \bibinfo{booktitle}{\emph{Proceedings of the 15th {Annual} {Conference} on
  {Information} technology education}}. \bibinfo{publisher}{ACM},
  \bibinfo{address}{Atlanta Georgia USA}, \bibinfo{pages}{39--44}.
\newblock
\showISBNx{978-1-4503-2686-5}
\urldef\tempurl%
\url{https://doi.org/10.1145/2656450.2656462}
\showDOI{\tempurl}


\bibitem[Trabelsi and Alketbi(2013)]%
        {trabelsi_using_2013}
\bibfield{author}{\bibinfo{person}{Zouheir Trabelsi} {and}
  \bibinfo{person}{Latifa Alketbi}.} \bibinfo{year}{2013}\natexlab{}.
\newblock \showarticletitle{Using network packet generators and snort rules for
  teaching denial of service attacks}. In \bibinfo{booktitle}{\emph{Proceedings
  of the 18th {ACM} conference on {Innovation} and technology in computer
  science education}}. \bibinfo{publisher}{ACM}, \bibinfo{address}{Canterbury
  England, UK}, \bibinfo{pages}{285--290}.
\newblock
\showISBNx{978-1-4503-2078-8}
\urldef\tempurl%
\url{https://doi.org/10.1145/2462476.2465580}
\showDOI{\tempurl}


\bibitem[Trabelsi and McCoey(2016)]%
        {trabelsi_ethical_2016}
\bibfield{author}{\bibinfo{person}{Zouheir Trabelsi} {and}
  \bibinfo{person}{Margaret McCoey}.} \bibinfo{year}{2016}\natexlab{}.
\newblock \showarticletitle{Ethical {Hacking} in {Information} {Security}
  {Curricula}}.
\newblock \bibinfo{journal}{\emph{International Journal of Information and
  Communication Technology Education}} \bibinfo{volume}{12},
  \bibinfo{number}{1} (\bibinfo{date}{Jan.} \bibinfo{year}{2016}),
  \bibinfo{pages}{1--10}.
\newblock
\showISSN{1550-1876, 1550-1337}
\urldef\tempurl%
\url{https://doi.org/10.4018/IJICTE.2016010101}
\showDOI{\tempurl}


\bibitem[Walshe and Simpson(2020)]%
        {walshe_empirical_2020}
\bibfield{author}{\bibinfo{person}{Thomas Walshe} {and} \bibinfo{person}{Andrew
  Simpson}.} \bibinfo{year}{2020}\natexlab{}.
\newblock \showarticletitle{An {Empirical} {Study} of {Bug} {Bounty}
  {Programs}}. In \bibinfo{booktitle}{\emph{2020 {IEEE} 2nd {International}
  {Workshop} on {Intelligent} {Bug} {Fixing} ({IBF})}}.
  \bibinfo{address}{London, ON, Canada}, \bibinfo{pages}{35--44}.
\newblock
\urldef\tempurl%
\url{https://doi.org/10.1109/IBF50092.2020.9034828}
\showDOI{\tempurl}


\bibitem[Weulen~Kranenbarg et~al\mbox{.}(2018)]%
        {weulen_kranenbarg_dont_2018}
\bibfield{author}{\bibinfo{person}{Marleen Weulen~Kranenbarg},
  \bibinfo{person}{Thomas~J. Holt}, {and} \bibinfo{person}{Jeroen van~der
  Ham}.} \bibinfo{year}{2018}\natexlab{}.
\newblock \showarticletitle{Don’t shoot the messenger! {A} criminological and
  computer science perspective on coordinated vulnerability disclosure}.
\newblock \bibinfo{journal}{\emph{Crime Science}} \bibinfo{volume}{7},
  \bibinfo{number}{1} (\bibinfo{date}{Dec.} \bibinfo{year}{2018}),
  \bibinfo{pages}{16}.
\newblock
\showISSN{2193-7680}
\urldef\tempurl%
\url{https://doi.org/10.1186/s40163-018-0090-8}
\showDOI{\tempurl}


\bibitem[Zetter(2001)]%
        {zetter_information_2001}
\bibfield{author}{\bibinfo{person}{Kim Zetter}.}
  \bibinfo{year}{2001}\natexlab{}.
\newblock \bibinfo{title}{Information {Security} {News}: {Three} {Minutes}
  {With} {Microsoft}'s {Scott} {Culp}}.
\newblock
\newblock
\urldef\tempurl%
\url{https://seclists.org/isn/2001/Oct/37}
\showURL{%
\tempurl}


\bibitem[Ķinis(2018)]%
        {kinis_responsible_2018}
\bibfield{author}{\bibinfo{person}{Uldis Ķinis}.}
  \bibinfo{year}{2018}\natexlab{}.
\newblock \showarticletitle{From {Responsible} {Disclosure} {Policy} ({RDP})
  towards {State} {Regulated} {Responsible} {Vulnerability} {Disclosure}
  {Procedure} ({RVDP}): {The} {Latvian} approach}.
\newblock \bibinfo{journal}{\emph{Computer Law \& Security Review}}
  \bibinfo{volume}{34}, \bibinfo{number}{3} (\bibinfo{date}{June}
  \bibinfo{year}{2018}), \bibinfo{pages}{508--522}.
\newblock
\showISSN{02673649}
\urldef\tempurl%
\url{https://doi.org/10.1016/j.clsr.2017.11.003}
\showDOI{\tempurl}


\end{thebibliography}


\end{document}